\documentclass[titlepage,11pt]{amsart}
\usepackage[margin=1in]{geometry}
\usepackage{amssymb, latexsym, cmll, stmaryrd,eqnarray}
\usepackage{mathtools}
\usepackage{xspace}
\usepackage{multirow}
\usepackage{amsaddr}
\usepackage{booktabs}
\usepackage{subcaption}
\usepackage{hyperref}
\usepackage{datetime}
\usepackage{lineno}

\calclayout

\usepackage{todonotes}

\usepackage{hyperref}
\makeatletter
\def\set@curr@file#1{%
  \begingroup
    \escapechar\m@ne
    \xdef\@curr@file{\expandafter\string\csname #1\endcsname}%
  \endgroup
}
\def\quote@name#1{"\quote@@name#1\@gobble""}
\def\quote@@name#1"{#1\quote@@name}
\def\unquote@name#1{\quote@@name#1\@gobble"}
\makeatother

\makeatletter
\def\set@curr@file#1{%
  \begingroup
    \escapechar\m@ne
    \xdef\@curr@file{\expandafter\string\csname #1\endcsname}%
  \endgroup
}
\def\quote@name#1{"\quote@@name#1\@gobble""}
\def\quote@@name#1"{#1\quote@@name}
\def\unquote@name#1{\quote@@name#1\@gobble"}
\makeatother

\newtheorem{lm}{Lemma}[section]
\newtheorem{thm}[lm]{Theorem}

\newtheorem{prp}[lm]{Proposition}
\newtheorem{cor}[lm]{Corollary}
\newtheorem{rmk}[lm]{Remark}
\newtheorem{ex}[lm]{Example}

\newtheorem{fact}[lm]{Fact}

\newcommand{\myqed}{\hfill$\Box$}

\newcounter{senumi}[section]
\newcounter{senumip}[section]
\newcounter{temp}[section]

\def\thesenumi{\thesection.\arabic{senumip}}
\def\p@senumip\thesenumip{\thesenumi}
    {\begin{list}%
        {(\thesenumi)}%
        {\usecounter{senumip}}
        \setcounter{senumip}{\value{temp}}
    }%
    {\setcounter{temp}{\value{senumip}}
     \end{list}}
\newcounter{penumi}[section]

\newcounter{ptemp}[section]

   {\begin{enumerate}%
        \setcounter{penumi}{\value{ptemp}}%
        \setcounter{enumi}{\value{ptemp}}%
    }%
    {\setcounter{ptemp}{\value{enumi}}
     \end{enumerate}}
\newcounter{ppenumi}[section]

\newcounter{pptemp}[section]

\def\theppenumi{\theptemp.\arabic{ppenumi}}
    {\begin{list}%
        {(\theppenumi)}%
        {\usecounter{ppenumi}\setlength{\rightmargin}{\leftmargin}}
        \setcounter{ppenumi}{\value{pptemp}}
    }%
    {\setcounter{pptemp}{\value{ppenumi}}
     \end{list}}

\newcommand{\m}[1]{{\uppercase {\mathbf{#1}}}}

\newcommand{\ceqv}[1]{\ensuremath{\operatorname{\textsc{{CEqv}}}
                                \ifthenelse{\equal{#1}{}}{}{\!\left( { #1} \right)}}}
\newcommand{\csat}[1]{\ensuremath{\operatorname{\textsc{{CSat}}}
                                \ifthenelse{\equal{#1}{}}{}{\!\left( { #1} \right)}}}
\newcommand{\scsat}[1]{\ensuremath{\operatorname{\textsc{\upshape{SCsat}}}
                                \ifthenelse{\equal{#1}{}}{}{\!\left( { #1} \right)}}}
\newcommand{\Csat}[1]{\ensuremath{\operatorname{\textsc{\upshape{Csat}}}
                                \ifthenelse{\equal{#1}{}}{}{\!\left( {\m #1} \right)}}}
\newcommand{\Ceqv}[1]{\ensuremath{\operatorname{\textsc{\upshape{Ceqv}}}
                                \ifthenelse{\equal{#1}{}}{}{\!\left( {\m #1} \right)}}}
\newcommand{\mcsat}[1]{\ensuremath{\operatorname{\textsc{\upshape{MCsat}}}
                                \ifthenelse{\equal{#1}{}}{}{\!\left( {\m #1} \right)}}}
\newcommand{\SCsat}[1]{\ensuremath{\operatorname{\textsc{\upshape{SCsat}}}
                                \ifthenelse{\equal{#1}{}}{}{\!\left( {\m #1} \right)}}}

\newcommand{\polsat}[1]{\ensuremath{\operatorname{\textsc{{PolSat}}}
                                \ifthenelse{\equal{#1}{}}{}{\!\left( {\m #1} \right)}}}
\newcommand{\poleqv}[1]{\ensuremath{\operatorname{\textsc{{PolEqv}}}
                                \ifthenelse{\equal{#1}{}}{}{\!\left( {\m #1} \right)}}}

\newcommand{\csp}[1]{\ensuremath{\operatorname{\textsc{\upshape{CSP}}}
                                \ifthenelse{\equal{#1}{}}{}{\!\left( {#1} \right)}}}

\newcommand{\npc}{\textsf{NP}-complete\xspace}

\newcommand{\ptime}{\textsf{PTIME}\xspace}
\newcommand{\expexp}{\textsf{2-EXPTIME}\xspace}

\newcommand{\set}[1]{{\left\{ {#1} \right\} }}
\newcommand{\ci}{\subseteq}

\newcommand{\card}[1]{\left| #1 \right|}

\renewcommand{\leq}{\leqslant}
\renewcommand{\geq}{\geqslant}

\renewcommand{\mapsto}{\longmapsto}

\newcommand{\jjoin}{\bigvee}
\newcommand{\mmeet}{\bigwedge}

\renewcommand{\o}[1]{\overline {#1}}
\newcounter{ttable}

\newcommand{\map}{\longrightarrow}

\newcommand{\h}[1]{\widehat{#1}}

\newcommand{\epsi}{\varepsilon}

\newcommand{\dpp}[1]{\m D[{#1}]}

\newcommand{\pand}{\textsf{AND}\xspace}
\newcommand{\por}{\textsf{OR}\xspace}
\newcommand{\pxor}{\textsf{XOR}\xspace}

\newcommand{\pcmod}[2]{\textsf{MOD}_{#1}^{#2}}

\newcommand{\cmod}{\textsf{MOD}\xspace}
\newcommand{\bb}{\mathsf{b}}
\newcommand{\ee}{\mathsf{e}}

\newcommand{\mytop}[2]{\genfrac{}{}{0pt}{}{#1}{#2}}

\newcommand{\zex}[1]{\mathbb{Z}\left[#1\right]}
\newcommand{\btyp}[1]{\left[#1\right]}

\newcommand{\zpq}{\zex{p,q}}
\newcommand{\zpx}{\mathbb{Z}_p[\o x]}
\newcommand{\zp}{\mathbb{Z}_p}
\newcommand{\zq}{\mathbb{Z}_q}

\newcommand{\cch}[2]{{\mathrm{CC}_{#1}{\left[{#2}\right]}}}
\newcommand{\cc}[1]{{\mathrm{CC}{\left[{#1}\right]}}}
\newcommand{\cll}{\ell}     
\newcommand{\rbit}{l}    
\newcommand{\kap}{p^*}

\newcommand{\alf}{\alpha}
\newcommand{\er}{\omega}
\newcommand{\err}{{(\er-1)}}
\newcommand{\es}{\varpi}
\newcommand{\ess}{{(\es-1)}}

\newcommand{\niu}{\nu}
\newcommand{\eh}{{h'}}

\newcommand{\zz}{{Z}}
\newcommand{\zzz}{\sigma}

\newcommand{\stack}[1]{D\left(#1\right)}
\newcommand{\subst}[3]{#1\left[#2/#3\right]}
\newcommand{\pre}[2]{#1^{-1}\left(#2\right)}
\newcommand{\pree}[2]{#1^{-1}{#2}}
\newcommand{\bal}[1]{\textsf{bal}\left(#1\right)}

\newcommand{\gam}{\gamma}
\newcommand{\gamon}{\gam_{h,m}}

\newcommand{\poly}[1]{poly #1}

\newcommand{\ethh}{Exponential Time Hypothesis\xspace}
\newcommand{\ethhh}{ETH\xspace}

\DeclareMathOperator{\polylog}{polylog}
\newcommand{\pr}[1]{\textrm{Pr}\left[ {#1} \right]}
\newcommand{\spike}[2]{\delta_{#1}^{#2}}
\newcommand{\dhr}{\dpp{p_0;p_1\cdot\ldots\cdot p_\er}}

\begin{document}

\title{Complexity of Modular Circuits}

\author{Pawe\l{} M. Idziak, \ \ Piotr Kawa\l{}ek}
\address{Department of Theoretical Computer Science, Jagiellonian University, Kraków, Poland}
\email{pawel.idziak@uj.edu.pl,  piotr.kawalek@doctoral.uj.edu.pl}

\author{Jacek Krzaczkowski}
\address{Department of Computer Science, Maria Curie-Sk\l{}odowska University, Lublin, Poland}
\email{krzacz@poczta.umcs.lublin.pl}

\thanks{The project is partially supported by Polish NCN Grant \# 2014/14/A/ST6/00138.}
\date{}

\begin{abstract}
We study how the complexity of modular circuits computing \pand depends on the depth of the circuits and the prime factorization of the modulus they use.
In particular our construction of subexponential circuits of depth 2 for \pand helps us to classify (modulo \ethh) modular circuits with respect to the complexity of their satisfiability.
We also study a precise correlation between this complexity and the sizes of modular circuits realizing \pand.
In particular we use the superlinear lower bound from \cite{LowBounds}
to check satisfiability of $CC^0$ circuits in probabilistic $2^{O(n/\epsi(n))}$ time,
where $\epsi$ is some extremely slowly increasing function.
%
Moreover we show that \pand can be computed by a polynomial size 
modular circuit of depth 2 (with $O(\log n)$ random bits)
providing a probabilistic computational model that can not be derandomized.

We apply our methods to determine (modulo \ethhh)
the complexity of solving equations over groups of symmetries of regular polygons
with an odd number of sides.
These groups form a paradigm for some of the remaining cases in characterizing
finite groups with respect to the complexity of their equation solving.
\end{abstract}


\maketitle

\clearpage
\section{Introduction}
\label{sec:intro}

Due to the pioneering work of Cook satisfiability of Boolean circuits is among the most celebrated problems in computer science.
Although the problem itself is \npc, it becomes solvable in \ptime when restricted to circuits of special kinds, like monotone circuits or circuits with linear gates only.
Here by linear gate we mean \pxor of unbounded fan-in.
Such a gate simply checks the parity of the sum of inputs.
This has been generalized to the gates $\pcmod{m}{A}$ that check if the sum of inputs,
taken modulo $m$ belongs to the set $A \ci \set{0,\ldots,m-1}$.
Note here that traditionally only the sets $A=\set{0}$
(or dually, only $A=\set{1,\dots,m-1}$)
are allowed.
We will however always consider generalized modular gates,
i.e. $\pcmod{m}{A}$ with arbitrary $A$ and multiple wires between gates (including input gates).
These gates are to be used to build modular circuits of bounded depth.
More precisely for depth $h$ and modulus $m$
by $\cch{h}{m}$ we mean a class of circuits built of gates $\pcmod{m}{A}$,
possibly with different $A$ for each gate.
In Section \ref{sec:cchm} we also discuss modular circuits
with possibly different moduli on different levels.
Thus a $\cc{m_1;\ldots;m_h}$-circuit admits only gates of the type $\pcmod{m_i}{A}$
on the $i$-th level.

Our results start with the following full characterization of parameters $h$ and $m$
for which satisfiability of $\cch{h}{m}$-circuits ($\cch{h}{m}$-SAT for short)
is in \ptime.
In what follows, for a positive integer $m$
by $\omega(m)$ we denote the number of different prime factors of $m$.
\begin{thm}
\label{poly}
Let $h$ and $m$ be positive integers.
Then under the assumption of \ethhh the problem of
satisfiability for $\cch{h}{m}$-circuits is in \ptime
iff $m=1$ or $\omega(m)=1$.
\end{thm}

Our nonpolynomial lower bounds are based on the construction of relatively small
$\cch{2}{m}$-circuits computing $\pand_n$,
i.e. of size $2^{O(\sqrt[\er]{n}\log n)}$, where $\er=\omega(m)$.
This construction improves the one of Barrington, Beigel and Rudich \cite{BarOR}
where $3$ levels were used.
From the papers \cite{BST90,ST06} we know that for $p$ being a prime,
$\cc{m;p^k}$-circuits expressing \pand
need at least $2^{\Omega(n)}$ gates.
But the general case of $\cch{2}{m}$ has been opened.

It is also worth to notice here that the expressive power of modular circuits with 2 levels
is also very sensitive to the sets $A$ used in $\pcmod{m}{A}$.
Indeed in \cite{caussinus} Caussinus shows the very same lower bound $2^{\Omega(n)}$ for \pand
if on the second level only the set $A=\set{1,\ldots,m-1}$ is allowed.

\medskip
Next we show that not only the width of the modulus $m$, i.e. $\omega(m)$,
but also the circuit depth may substantially contribute to reduce the size of
$\cch{h}{m}$-circuits realizing \pand.
Surprisingly also the number $\varpi(m)$ of large prime factors of $m$ plays some role.
By a large prime divisor of $m$ we mean each one that is at least $\omega(m)$.

\begin{thm}
\label{thm:omegabar}
For $h\geq 3$ and a positive integer $m$ with $\er=\omega(m)\geq 2$ and $\es = \varpi(m)$
there are $\cch{h}{m}$-circuits of size $2^{O({n}^{1/(\err(h-2)+\es)}\log n)}$,
computing $n$-ary \pand.
\end{thm}

Although the only known lower bound
for the size of modular circuits computing \pand
is slightly better than linear (see \cite{LowBounds}),
Barrington, Straubing and Thérien \cite{BST90} conjectured that it has to be exponential.
In fact, after the paper \cite{BarOR}, the bound $2^{\Omega(n^\delta)}$ with some $\delta>0$
is a popular belief.
In contrast to this conjecture there are constructions \cite{hansen08,hansen-koucky09}
of (quasi)polynomial size probabilistic modular circuits computing \pand.
The construction in \cite{hansen08} is of quasipolynomial size and uses $\polylog(n)$ random bits.
The one from \cite{hansen-koucky09} fixes the depth to be constant (but a substantial one),
reduces the size to be polynomial
and cuts down the number of random bits to $O(\log n)$.
Our techniques applied in the proof of Theorem \ref{thm:omegabar}
have proved to be useful also in this probabilistic setting.
First, while keeping only $O(\log n)$ random bits,
we reduce the depth of the circuits realizing \pand to be only $2$.
Moreover our construction is more transparent, as it makes no use of expanders or universal hashing functions.

\begin{thm}
\label{thm:random-and}
For the modulus $m$ with $\omega(m)\geq 2$ the $n$-ary \pand functions
can be realized by $\cch{2}{m}$-circuits of polynomial size with $O(\log n)$ random bits.
In fact the realization is done by $\cc{p,q}$-circuits, where $p,q$ are different primes.
\end{thm}

Again, the mentioned lower bound $2^{\Omega(n)}$ 
for the size of deterministic $\cc{p,q}$-circuits computing \pand,
blocks any derandomization here.
Thus to confirm the suggestion made in \cite{hansen-koucky09}
that \pand can be computed by small ${\rm CC}^0$ circuits
one would need to increase the depth.

\medskip

We also show how superpolynomial lower bounds for sizes of modular circuits computing \pand
would give rise to subexponential algorithms checking satisfiability (or equivalence) of such circuits. In fact this connection, as well as the converse one,
(due to their technicality) is presented only in Section \ref{sec:algo},
in particular in  Theorem \ref{thm:algo}.
As a result of these considerations we obtain (see Theorem \ref{thm:cc-vs-ac}) an upper bound for satisfiability of $CC^0$-circuits that is asymptotically lower than the one \ethhh permits for $AC^0$-circuits.

\medskip
Our methods proved themselves to be powerful enough to be applied in some other contexts.
In particular in Theorem \ref{thm:dm} we give a characterization of dihedral groups $\m D_{2k+1}$,
i.e. groups of symmetries of regular polygon with odd number of sides,
for which the problem of solving equation is tractable.
In fact for odd $m$ we show that this happens only if $\omega(m)=1$, or \ethhh fails.
This result partially fills the small gap that remains, after the paper \cite{ikkw},
in characterizing finite groups with polynomial time algorithms for solving equations.

Another, in fact pretty similar, application of our methods is done
for satisfiability of multivalued circuits, as defined in \cite{ik:lics18}.

\section{Shallow or narrow may apply}
\label{may-apply}

In this section we analyze the expressive power of $\cch{h}{m}$-circuits
with $h=1$ or $\omega(m)=1$.
We start with stating that in such realm $\cch{h}{m}$-circuits can compute \pand only of bounded arity.
Although one can find proofs of very similar statements in the literature (e.g \cite{BST90,bt}),
we have decided to sketch the proof in Section \ref{sec:easy}.

\begin{prp}
\label{shallow-narrow-and}
For positive integers $m,h,k$ and a prime  $p$,
the arity of \pand \ computable by
\begin{itemize}
  \item $\cch{1}{m}$-circuits is bounded by $m-1$,
  \item $\cch{h}{p^k}$-circuits is bounded by a constant depending only on $h$ and $m=p^k$.
  \hfill\myqed
\end{itemize}
\end{prp}

From the above bound we can infer one implication in Theorem \ref{poly}.
To do this an easy observation is required, that we can simulate $\cch{h}{m}$ circuit with some inputs fixed to be constant, just by slightly modifying the structure of the circuit,
without inflating its size.
For this reason, in the following, we simply allow some inputs to be constant

\begin{cor}
\label{cor-shallow-narrow}
Satisfiability of $\cch{h}{m}$-circuits is in \ptime
whenever $h=1$ or $\omega(m)=1$.
\end{cor}

\begin{proof}
The only property of $\cch{1}{m}$ and $\cch{h}{p^k}$ circuits
we are going to use is that they have a bound, say $c$, for the arity of \pand they can express.
This bound allows us to reduce our search for an $n$-tuple $a$ in the large set $\set{0,1}^n$ satisfying the circuit $\Gamma$ to a smaller subset of size at most $n^c$
containing only tuples with at most $c$ ones.
Indeed, let $a$ be a satisfying tuple with minimal possible $\card{\pre a 1}$.
By this minimality we know that $\Gamma$ with all the inputs with indices outside $\pre a 1$
set to $0$ behaves like the $\card{\pre a 1}$-ary \pand.
Thus $\card{\pre a 1} \leq c$,
so that the tuple $a$ has at most $c$ ones, as claimed.
\end{proof}

The function $\pand_n$ is an example of a nonconstant extremely unbalanced boolean function,
i.e, one value is taken exactly once.
Our next goal is to show that shallow or narrow modular circuits can compute
functions with rather balanced piles, i.e. preimages $\pre f 0$ and $\pre f 1$.
Formally the balance of the $n$-ary boolean function $f$ is defined to be
\[
\bal{f} = 1- \frac{\card{\card{\pre f 0}-\card{\pre f 1}}}{2^n}.
\]
Constant functions have balance $0$.
The functions $\pand_n$ and $\por_n$ have the smallest possible non-zero balance $2^{1-n}$.
In fact each $n$-ary function $f$ with one element smaller pile, denoted later $\stack f$, has balance $2^{1-n}$ and is to be called a spike.
An obvious calculation shows that $\card{\stack f} = 2^{n-1}\bal f$.

\begin{rmk}
\label{rmk-reducing-stack}
Each nonconstant $n$-ary boolean function $f$
can be turned to be $(n-k)$-ary spike
by fixing its $k \leq \log\card{\stack f}$ variables to be constant from $\set{0,1}$.
\end{rmk}

\begin{proof}
To fix the notation we use the symbol $\subst{f}{x_i}{c}$ for the function
obtained from $f$ by fixing the variable $x_i$ to be $c\in\set{0,1}$.

Now, as long as $\card{\stack f}\geq 2$
we iteratively reduce the size of $\stack f$ 
at least twice,
by fixing the value of one of the variables without making $f$ constant.
Thus we start with picking $a^0,a^1 \in \stack f = \pre f b$
and a coordinate $i$ so that $a^0_i=0$ and $a^1_i=1$.
%
Obviously $\card{\pre f b} = \card{\pre{\subst{f}{x_i}{0}}{b}}+\card{\pre{\subst{f}{x_i}{1}}{b}}$
and we pick $c\in\set{0,1}$ so that
$\card{\pre{\subst{f}{x_i}{c}}{b}}\leq\card{\pre{\subst{f}{x_i}{1-c}}{b}}$.
Thus we have $\card{\stack{\subst{f}{x_i}{c}}} \leq \card{\stack{f}}/2$,
as required.
To see that $\subst{f}{x_i}{c}$ is not constant, note that $a^c$,
with its $i$-th coordinate removed, belongs to $\stack{\subst{f}{x_i}{c}}$
so that $1 \leq \card{\stack{\subst{f}{x_i}{c}}} \leq \card{\stack{f}}/2 <2^n/2=2^{n-1}$.
\end{proof}

We note here that all spikes (of the same arity) are interdefinable.
Indeed, if $\spike{\o a}{\epsi}$ denotes the spike which takes the value $\epsi\in\set{0,1}$
only on the tuple $\o a \in \set{0,1}^n$ then
\begin{itemize}
  \item $\spike{\o a}{1-\epsi}(\o x)=1-\spike{\o a}{\epsi}(\o x)$,
  \item $\spike{\o b}{\epsi}(x_1,\ldots,x_n)=\spike{\o a}{\epsi}(x_1+a_1-b_1,\ldots,x_n+a_n-b_n)$.
\end{itemize}
This interdefinability can be realized by modifying only the sets $A$ in the gates $\pcmod{m}{A}$
on the last and/or first level,
so that the sizes of the corresponding circuits remain unchanged.

Now, if the arity of spikes computable by $\cch{h}{m}$-circuits is bounded,
like in Proposition \ref{shallow-narrow-and},
we use Remark \ref{rmk-reducing-stack} to turn the circuit into the one computing
a spike with arity at least $n-\log\card{\stack f} = n - (n-1)\log\bal f$.
This gives the following lower bound on the balance of $\cch{h}{m}$-circuits
independently of its arity and size.

\begin{cor}
\label{cor:bal1}
For $h=1$ or $\omega(m)=1$ the balance of non-constant functions computable by $\cch{h}{m}$-circuits
is at least $2^{1-c}$, where $c$ bounds the arity of $\cch{h}{m}$-computable conjunctions.
\end{cor}

From Corollary \ref{cor:bal1} we immediately get the following.

\begin{cor}
\label{cor:lan-bal1}
Let $L\ci\set{0,1}^*$ be a language recognizable by $\cch{h}{m}$ circuits,
where $h=1$ or $\omega(m)=1$.
Then the number of words in $L$ of the length $n$ is either $0$ or is rather large,
i.e. at least $2^{n-c}$,
where $c$ bounds the arity of $\cch{h}{m}$-computable conjunctions.
\end{cor}
\section{Deep or wide need not apply}
\label{sec:cc2}

In this section we will show the converse to Corollary \ref{cor-shallow-narrow}
but under the assumption of the \ethh.
As we have already noted this is done by constructing conjunction of subexponential size.
The next Proposition formulates this fact in more details.

\begin{prp}
\label{prp:force-narrow}
For a positive integer $m$ with exactly $\er$ different prime divisors 
we have:
\begin{enumerate}
  \item\label{force-narrow-cnf}
        for each 3-CNF-SAT formula $\Phi$ with $\cll$ clauses
        there is a $\cch{2}{m}$ circuit of size at most $2^{O(\sqrt[\er]{\cll}\log \cll)}$
        representing $\Phi$,
  \item \label{force-narrow-and}
        in particular unbounded fan-in \pand \ can be computed by $\cch{2}{m}$ circuits of size
        $2^{O(\sqrt[\er]{n}\log n)}$,
        where $n$ is the number of variables (input gates).
\end{enumerate}
The above bounds on the circuits size also bound the time needed to obtain them.
\end{prp}

Combining this Proposition with \ethh (and Sparsification Lemma) we immediately get the following Corollary.

\begin{cor}
\label{eth-nonpoly}
If $h\geq 2$ and $\omega(m)\geq 2$
then satisfiability for $\cch{h}{m}$-circuits is not in \ptime,
unless \ethhh fails.
\end{cor}

As we have mentioned in the Introduction our proof of Proposition \ref{prp:force-narrow}
is modelled after the idea of Barrington, Beigel and Rudich \cite{BarOR}
where the $\pand_n$ had been shown to be computable by modular circuits of the same subexponential size as described in Proposition \ref{prp:force-narrow}(\ref{force-narrow-and}) but on $3$ levels.
In squeezing this to $2$ levels we need the concepts of $\zpq$-expressions
and the circuits realizing them.

In the papers \cite{ikk:mfcs18, ikk:lics20} we have been studied action of the group
$\zp$ on the group $\zq$ via the function $\bb : \zp \map \zq$
defined by $\bb(0)=0$ and $\bb(x)=1$ for all other $x\in \zp$.
With the help of this action we define $\zpq$-expression to be 
the $n$-ary expression over the variables $\o x =(x_1,\ldots, x_n)$:
\[
t(\o x) =
\sum_{\mytop{\beta\in Z_p^n}{c\in Z_p}}
\alpha_{\beta,c}\cdot \bb\left(\sum_{i=1}^n \beta_i x_i +c\right),
\]
where the $\alpha_{\beta,c} \in Z_q$
while $\beta=(\beta_1,\ldots,\beta_n)\in Z_p^n$ and $c \in Z_p$,
the outer sum and the multiplications by the $\alpha_{\beta,c}$'s are taken modulo $q$,
while the inner sum and the multiplications by the $\beta_i$'s are taken modulo $p$.
Obviously the $\zpq$-expression $t(\o x)$ is determined by the sequence
$\left\langle  \alpha_{\beta,c} : {\beta\in Z_p^n},{c\in Z_p}\right\rangle$
of coefficients from $\mathbb{Z}_q$.
This sequence may have the exponential size $p^{n+1}$.
However only the nonzero  $\alpha_{\beta,c}$'s contribute to the length of $t(\o x)$
and consequently to the size of a circuit that models $t(\o x)$.
In fact, if $t(\o x)$ returns always boolean values on boolean inputs $\o x$,
$t(\o x)$ may be realized by a circuit, called $\Gamma(t)$,
of size $1+\card{L(t)}$,
where $L(t)= \set{(\beta,c) \in Z_q^n \times Z_q : \alpha_{\beta,c} \neq 0}$.
Indeed, the subexpression $\bb\left(\sum_{i=1}^n \beta_i x_i +c\right)$
can be realized by a single $\cmod_{p}^{\mathbb{Z}_{p}-\set{-c}}$ gate
(denoted $\Gamma_{\beta,c}(t)$),
and then combining the outputs of all the $\Gamma_{\beta,c}(t)$
(with $(\beta,c)$ ranging over $L(t)$)
by the $\cmod_q$-like gate.
For this reason by the size of the circuit $\Gamma(t)$,
as well as of the $\zpq$-expression $t(\o x)$, we simply mean $1+\card{L(t)}$.

In our further consideration we will also use the bunch
$\Theta(t) = \set{\Gamma_{\beta,c}}_{(\beta,c)\in L(t)}$ of the above gates
with $\card{L(t)}$ outputs.
These outputs are going to be treated as a single bundle
(without ordering, but with copying the output of each $\Gamma_{\beta,c}(t)$ the corresponding, i.e. $\alpha_{\beta,c}$, number of times)
as they always will be used as inputs to other $\cmod$-gates, so that they will be summed up first.
To keep track of the modulus used to sum up this bundle we will say that the bundle is of type $q$ and that the bunch $\Theta(t)$ is of type $\btyp{p,q}$.

\bigskip
The importance of the $\zpq$-expressions lies in the next Fact that has been originally shown
is \cite{ikk:mfcs18} as Lemma 3.1
(but with $\bb(x)$ replaced by $\h\bb(x)= 1-\bb(1-x)$).

\begin{fact}
\label{trzy-jeden}
With two different primes $p,q$ we can represent every $n$-ary function
$g: Z_p^n \map Z_q$ by a $\zpq$-expression of length and size bounded by $2^{O(n)}$.
\hfill\myqed
\end{fact}

The next fact is borrowed from \cite{BarOR},
but we include its more transparent proof in Section \ref{sec:easy}.

\begin{fact}
\label{bar-poly}
Let $p$ be a prime and $k\geq 1$ be an integer.
Then there is a polynomial $w(\o x) \in \zpx$ of degree at most $p^k-1$,
such that for $\o x \in \set{0,1}^n$ we have
\[
w(\o x) =
\left\{
\begin{array}{ll}
0, &\mbox{if \ $\card{\pre {\o x}{0}}  \equiv 0$ modulo $p^k$,}\\ 
1, &\mbox{else.}
\end{array}
\right.
\]
\end{fact}

With the help of Facts \ref{trzy-jeden} and \ref{bar-poly}
we can represent 3-CNF formulas by a relatively short $\zpq$-expressions
in the following sense:

\begin{lm}
\label{lm:pseudo-and}
Let $p,q$ be two different primes and $\niu \geq 1$ be an integer.
Then for each 3-CNF-SAT formula $\Phi(\o x)$
with $n$ variables $\o x = (x_1,\ldots, x_n)$
and $\cll$ clauses
there is a $\zpq$-expression $t^\Phi_{p,q}(\o x)$
of size at most $2^{O(q^\niu \cdot \log \cll)}$
such that for all $\o a \in \set{0,1}^n$ we have
\[
t^\Phi_{p,q}(\o a) =
\left\{
\begin{array}{ll}
0, &\mbox{if the number of unsatisfied (by $\o a$) clauses in $\Phi$ is divisible by $q^\niu$}\\
1, &\mbox{else.}
\end{array}
\right.
\]
\end{lm}

\begin{proof}
To fix our notation let
$\Phi(\o x) = \mmeet_{i=1}^\cll C_i$ be a 3-CNF formula
with the clauses $C_i= C_i(z^1_i,z^2_i,z^3_i)$.
Fact \ref{bar-poly} supplies us with an $\cll$-ary polynomial
$w(c_1,\ldots, c_\cll) \in GF(q)[\o c]$ of degree at most $q^\niu-1$.
We want to feed up the polynomial $w$ by substituting $C_i(z^1_i,z^2_i,z^3_i)$ for the variable $c_i$ to get a total function $w^* : Z_p^n \map Z_q$.
In order to do that we first extend each clause $C_i$ to be a total function
$Z_p^3 \map Z_q$ (instead of $\set{0,1}^3 \map \set{0,1}$)
by putting arbitrary values on the set $Z_p^3 - \set{0,1}^3$.
Now the function
\[
w^*(\o z) = w\left(C_1(z^1_1,z^2_1,z^3_1),\ldots,C_\cll(z^1_\cll,z^2_\cll,z^3_\cll)\right)
\]
behaves on the boolean values exactly as we need, i.e. for $\o a \in \set{0,1}^n$ we have
\[
w^*(\o a) =
\left\{
\begin{array}{ll}
0, &\mbox{if the number of unsatisfied (by $\o a$) clauses in $\Phi$ is divisible by $q^\niu$}\\
1, &\mbox{else.}
\end{array}
\right.
\]
All we need is to turn $w^*$ into a relatively short $\zpq$-expression.
Instead of applying Fact \ref{trzy-jeden} directly to $w^*$
we will do it for each its monomial separately.
Note that the monomials of $w$, after our substitution of the $C_i$'s for $c_i$'s
have the form
\[
C_{i_1}(z^1_{i_1},z^2_{i_1},z^3_{i_1}) \cdot \ldots \cdot C_{i_s}(z^1_{i_s},z^2_{i_s},z^3_{i_s})
\mbox{ \ \ \  with \ \ \ } s<q^\niu,
\]
so that there are at most $3q^\niu$ variables involved into each such ``monomial''.
Because of that, Fact \ref{trzy-jeden} allows us to represent each summand in $w^*$
by a $\zpq$-expression of size $O(2^{cq^\niu})$.
Since $\cll^{q^\niu}$ bounds the number of monomials of degree at most $q^\niu-1$
it also bounds the number of summands in $w^*$,
so that we end up with the bound
$O\left( 2^{cq^\niu} \cdot \cll^{q^\niu} \right)\leq 2^{ O\left({q^\niu \log \cll} \right)}$
for our $\zpq$-expression representing $w^*$.
\end{proof}

Our Claim shows also that for $p,q,\niu$ as above
we also have a relatively short $\zpq$-expression $t_{p,q}(x_1,\ldots,x_n)$
that behaves almost like an $n$-ary \pand.
That is, its size is at most $2^{O({q^\niu \cdot \log n})}$
and for all $\o a \in \set{0,1}^n$ we have
\begin{eqnarray}
\label{pq-exp}
t_{p,q}(\o a)&=&
\left\{
\begin{array}{ll}
0, &\mbox{if the number of zeros among the $a_i$'s is divisible by $q^\niu$},\\
1, &\mbox{else.}
\end{array}
\right.
\end{eqnarray}
We will also use the symbols
$\Gamma_{p,q},\Gamma^\Phi_{p,q}$ to denote the circuits
$\Gamma(t_{p,q}), \Gamma(t^\Phi_{p,q})$
computing the $\zpq$-expressions $t_{p,q}$ and $t^\Phi_{p,q}$, respectively.
Also the symbols $\Theta_{p,q},\Theta^\Phi_{p,q}$
will be used to denote the bunch of initial $\cmod_p$-gates in the circuits $\Gamma_{p,q},\Gamma^\Phi_{p,q}$.

\medskip
Now we are ready to prove  Proposition \ref{prp:force-narrow}.
\begin{proof}
To start we let $m=p_1^{\alf_1}\cdot\ldots p_\er^{\alf_\er}$ be the prime decomposition of $m$.
Each of the groups $\mathbb{Z}_{p_j}$'s can be identified with a subgroup of $\mathbb{Z}_m$
generated by $\frac{m}{p_j}$,
by simply sending $z$ to $\frac{m}{p_j} \cdot z$.
After such identification we know that the sum
$\sum_{j=1}^{\er} \frac{m}{p_j} \cdot \mathbb{Z}_{p_j}$ is in fact a direct sum,
so that each element of this sum has a unique decomposition.

To construct a $\cch{2}{m}$ circuit computing the 3-CNF formula $\Phi$ with $\cll$ clauses
we first fix integers $\niu_1,\ldots,\niu_\er$ to satisfy
$p_j^{\niu_j-1} \leq \sqrt[\er]{\cll} < p_j^{\niu_j}$.
Also, for convenience we identify the index $0$ with $\er$
so that we can refer to the indices of the primes $p_j$'s cyclically.
Now, for each $j=1,\ldots,\er$, \ Lemma \ref{lm:pseudo-and}
supplies us with a $\zex{p_{j-1},p_{j}}$-expression
$t^\Phi_j(\o x)$ of the length at most $O(2^{c_j\cdot \sqrt[\er]{\cll} \cdot \log \cll})$
so that for $\o a \in \set{0,1}^n$ we have
\[
t^\Phi_j(\o a) =
\left\{
\begin{array}{ll}
0, &\mbox{if the number of unsatisfied (by $\o a$) clauses in $\Phi$ is divisible by $p_j^{\niu_j}$}\\
1, &\mbox{else.}
\end{array}
\right.
\]
Our identification of the direct sum $\bigoplus_{j=1}^\er \frac{m}{p_j} \cdot \mathbb{Z}_{p_j}$ with a subgroup of $\mathbb{Z}_m$ allows us to sum up (modulo $m$) all the $t^\Phi_j$ to get
\begin{equation}
\label{tfi}
T^\Phi(\o a) = \sum_{j=1}^{\er} \frac{m}{p_j} \cdot t^\Phi_j(\o a).
\end{equation}
We argue now that  for $\o a \in \set{0,1}^n$
\[
\mbox{$T^\Phi(\o a) = 0$ \ \ iff \ \ $\Phi$ is satisfied by $\o a$}.
\]
Indeed, to see the `if' direction note that the number $\cll_0$ of unsatisfied (by $\o a$) clauses
is zero so that Lemma \ref{lm:pseudo-and} gives that each of the $t^\Phi_j(\o a)$'s, and consequently the sum $T^\Phi(\o a)$, is zero.
Conversely, if $\o a$ does not satisfy $\Phi$ then $1\leq \cll_0$,
which together with $\ell_0 \leq \cll < p_1^{\niu_1}\cdot\ldots\cdot p_\er^{\niu_\er}$ gives that at least one of the $p_j^{\niu_j}$'s does not divide $\cll_0$.
Thus for this $j$ the summand $\frac{m}{p_j} \cdot t^\Phi_j(\o a)$ is non-zero
and -- by the unique decomposition -- the entire sum $T^\Phi(\o a) \neq 0$.

Now, the circuit required in Proposition \ref{prp:force-narrow}\eqref{force-narrow-cnf}
is not supposed to calculate separately each of the $\Gamma(t^\Phi_j)$'s
by summing up the subexpressions
$\bb\left(\sum_{i=1}^n \beta_i x_i +c\right)$ of $t^\Phi_j$.
Instead, each such subexpression is calculated by the gate $\Gamma_{\beta,c}(t^\Phi_j)$
and then sent to $\cmod_m^{\set{0}}$-gate  $\left(\frac{m}{p_j}\cdot\alpha_{\beta,c}\right)$-times.
Due to the properties of $T^\Phi$, this last gate,
after collecting all the bundles $\Theta(t^\Phi_j)$'s,
calculates the boolean value of $\Phi(\o a)$.
Moreover the entire circuit consists of $1+\sum_{j=1}^r \card{L(t_j)}$ gates:
\begin{itemize}
  \item the final gate $\cmod_m^{\set{0}}$,
  \item the gates $\Gamma_{\beta,c}(t^\Phi_j)$ of the form $\cmod_{p_j}^{\mathbb{Z}_{p_j}-\set{-c}}$, one for each $(\beta, c) \in L(t_j)$.
\end{itemize}
From Lemma \ref{lm:pseudo-and} we know that the sizes of the $t_j$'s
(and therefore of the $\Theta(t^\Phi_j)$'s)
can be uniformly bounded by $O(2^{c\sqrt[\er]{\cll}\log \cll})$.
Thus this also bounds the size of the circuit.
\end{proof}

Note here that in our construction of subexponential size $\cch{2}{m}$-circuit computing \pand
the final gate is $\cmod_m^{\set{0}}$.
This contrasts the result of Caussinus \cite{caussinus}
where the lower bound $2^{\Omega(n)}$ is shown if the final gate is
$\cmod_m^{\set{1,\ldots,m-1}}$.

\section{Making the circuits smaller}
\label{sec:cchm}

We start with observing that composing $\cch{2}{m}$ circuits
by using 2 separate groups of 2 levels we can keep the size
$2^{O(k^{1/\er}\log k)}$ to compute $\pand_{k^2}$.
Indeed we can simply feed the $k$ inputs of the last two levels computing $\pand_k$
by the outputs of $k$-ary independent conjunctions built on the 2 starting levels.
Repeating this recursively $\lfloor h/2 \rfloor$-many times on $h$ levels
we get the following Proposition.

\begin{prp}
\label{prp:r-and-h-pol}
For $h\geq 2$ and a positive integer $m$ with $\er=\omega(m)\geq 2$
there are $\cch{h}{m}$-circuits of size
$2^{O({n}^{1/(\er\lfloor h/2 \rfloor)}\log n)})$,
computing $n$-ary \pand.
\end{prp}

Our next step is to use both the depth $h$ of the circuit
and the width $\omega(m)$ of the modulus to make our $\cch{h}{m}$-circuits
for \pand much smaller.
But before doing that we warm up with the following easy observation.
The idea of its proof has been already
explored in \cite{ikk:lics20,weiss:icalp20,ikkw}.
\begin{prp}
\label{prp:single-primes}
For $h\geq 2$ and a sequence of alternating primes $p_1 \neq p_2 \neq p_3 \neq \ldots \neq p_h$
there are $\cc{p_1;\ldots;p_h}$-circuits of size
$2^{O({n}^{1/(h-1)})}$,
computing $n$-ary \pand.
\end{prp}

\begin{proof}
Obviously we may assume that $n=k^{h-1}$ for some $k$.
For each $j=1,\ldots,h-1$ Fact \ref{trzy-jeden} supplies us with a $k$-ary $\mathbb{Z}[p_j,p_{j+1}]$-expression $C_j$ of size $2^{O(k)}$ that on
$\o a\in\set{0,1}^k\ci \mathbb{Z}_{p_j}^k$ behaves as $\pand_k$.
On the starting level of our circuit we group $n=k^{h-1}$ inputs
into $n/k$ groups of $k$ inputs each.
Then each group is passed through the bunch $\Theta(C_1)$
so that we end up with $n/k$ bundles $B_i$.
Note that if $B_i$ was passed through
$\pcmod{p_2}{\set{1}}$ gate we would get the conjunction of $k$ inputs of $B_i$.
Instead we again group the bundles $B_1,\ldots,B_{n/k}$ into $n/k^2$ groups with $k$ bundles each
and pass each such a group through the bunch $\Theta(C_2)$.
Again, the sum of each of the $n/k^2$ resulting bundle (modulo $p_3$)
coincide with \pand of ${k^2}$ on the initial inputs that fall into that bundle.
After repeating this $h-1$ times we end up with a single bundle of type $p_{h}$.
At this point we actually use $\pcmod{p_h}{\set{1}}$ gate to sum this bundle up and get
\pand of all the inputs.

It should be clear that the size of the entire circuit
is bounded by $2^{O(k)} = 2^{O({n}^{1/(h-1)})}$.
\end{proof}

Now we are in a position to prove Theorem \ref{thm:omegabar}.
However we will start with its slightly weaker version.

\begin{prp}
\label{prp:omega-omegabar}
For $h\geq 3$ and a positive integer $m$ with $\er=\omega(m)\geq 2$ and $\es = \varpi(m)$
there are $\cch{h}{m}$-circuits of size $2^{O({n}^{1/(\err(h-2)+\ess)}\log n)}$,
computing $n$-ary \pand.
\end{prp}

\begin{proof}
As in the proof of Proposition \ref{prp:force-narrow}
we start with the prime decomposition
$m=p_1^{\alf_1}\cdot\ldots p_\er^{\alf_\er}$
and assume that
$p_1 > \ldots > p_\es \geq \er >p_{\es+1}>\ldots > p_\er$.
Moreover, without loss of generality we assume that
$n=k^{\err(h-2)+\ess}$ for some integer $k$
and put $k_\er=\err k^{\er-1}$ and $k_\es=\err k^{\es-1}$.
Finally we pick integers
\[
\begin{array}{lcccll}
\nu_1,\ldots,\nu_\er &      \mbox{satisfying} &p_j^{\nu_j-1}   \leq k_\er^{1/\err} < p_j^{\nu_j},%
&\mbox{so that} & \prod_{j\neq i} p_j^{\nu_j} > k_\er, &\mbox{for $i=1,\dots,\er$,}
\\
\o\nu_1,\ldots,\o\nu_\es &  \mbox{satisfying} &p_j^{\o\nu_j-1} \leq k_\es^{1/\ess}< p_j^{\o\nu_j},%
&\mbox{so that} & \prod_{j\neq i} p_j^{\o\nu_j} > k_\es,  &\mbox{for $i=1,\dots,\es$.}
\end{array}
\]
Also for two different prime divisors $p,q$ of $m$
we modify $k_\er$-ary and $k_\es$-ary $\zpq$-expressions of the form $t_{pq}$ that satisfy (\ref{pq-exp}) to $t'_{pq} =1-t_{pq}$ with the arity that later should be clear from the context.
Note here that, except their arities, the $t_{p_ip_j}$'s depend not only on the primes $p_i, p_j$
but also on the integers $\nu_j$ (or $\o\nu_j$, whatever applies).

By $\Gamma'_{pq}$ and $\Theta'_{pq}$ we denote the circuit $\Gamma(t'_{pq})$
and the bunch $\Theta(t'_{pq})$ of type $[p,q]$, respectively.

Note that for fixed $p_i$ and $z_1,\ldots,z_{k_\er}\in \set{0,1}$ we have
\begin{equation}
\label{eqn:and}
\pand\set{t'_{p_ip_j}(z_1,\ldots,z_{k_\er}) : j\neq i} = \pand\set{z_1,\ldots,z_{k_\er}}.
\end{equation}
Indeed, Lemma \ref{lm:pseudo-and} assure us that the left hand side in the above display is $1$
if and only if for all $j\neq i$ the number of zeros among the $z$'s is divisible by $p_j^{\niu_j}$.
This in turn  means that the number of zeros among the $z$'s
is divisible by $\prod_{j\neq i} p_j^{\niu_j} > \prod_{j\neq i} k_\er^{1/\err} = k_\er$.
But there are only $k_\er$ places for such zeros so that there are no zeros among the $z$'s at all.

\medskip
Now for each $\eh=0,1,2,\ldots, h-2$ we recursively built a circuit $\nabla_\eh$ of depth $\eh$
\begin{enumerate}
  \item[(i)]
    with $n$ inputs $x_1,\ldots, x_n$,
    (repeated $\er\err$ times by $\nabla_0$)
  \item[(ii)]
    and with $b_\eh = \er\err\cdot n/k^{\err\eh} = \er\err \cdot k^{\err(h-2-\eh)+\ess}$
    bundles of outputs.
\end{enumerate}
For $\eh >0$ each bundle mentioned in (ii) is the result of some bunch of the form $\Theta_{pq}$.
Thus each bundle has one of the types $p_1,\ldots,p_\er$ and all the bundles are evenly divided into these types so that
\begin{enumerate}
  \item[(iii)]
    there are $b_\eh/\er = \err \cdot k^{\err(h-2-\eh)+\err}$ bundles of each type.
\end{enumerate}
Moreover enlarging $\nabla_\eh$ to $\nabla_{\eh+1}$ we will keep the following properties:
\begin{enumerate}
  \item[(iv)]
    summing up (modulo $q$) a bundle $B$ of type $q$ (to get $s_B(\o x)$)
    only the boolean values $0$ or $1$ may appear,
  \item[(v)]
    the conjunction of all $b_\eh$ values $s_B(\o x)$
    (i.e. with $B$ ranging over all bundles produced by $\nabla_\eh$)
    coincides with $\pand(x_1,\ldots,x_n)$.
\end{enumerate}
We start with artificially adding level $0$ just to multiply variables
so that it does not contribute to the depth of our circuits.
In fact this starting circuit $\nabla_0$ (of depth $0$)
takes $n$ inputs $x_1,\ldots,x_n$
and makes $b_0=\er\err \cdot n$ bundles,
each of which consisting of one typed variable,
i.e. each variable $x_i$ is repeated $\er-1$ times in each type.
It should be (more than) obvious that (i)-(v) hold.

Now to go from $\nabla_\eh$ to $\nabla_{\eh+1}$
we first group $\frac{b_\eh}{\er}$ bundles of a given type,
say $p$, into $\frac{b_\eh}{\er k_\er}$ groups of size $k_\er$
(i.e. each such a group consists of $k_\er$ bundles of type $p$).
Next, all $k_\er$ bundles in one group are passed through $\er-1$ bunches $\Theta'_{pq}$,
one for each $q\neq p$,
to produce $\er-1$ bundles, again one for each type $q\neq p$.
Thus $b_\eh$ bundles (that go to the gates on level $\eh+1$)
are replaced by $b_{\eh+1}=\err\frac{b_\eh}{k_\er}=\frac{b_\eh}{k^{\er-1}}$ new bundles,
as required in (i)-(iii).
To pass the $k_\er$-element group $B_1,\ldots,B_{k_\er}$ of bundles
through the bunch $\Theta'_{pq}$ of gates
we inflate each single input (say the $s$-th one) of $\Theta'_{pq}$ into the number of outputs in $B_s$ so that in fact $\Theta'_{pq}$ is fed by $s_{B_1},\ldots,s_{B_{k_\er}}$.

To see (iv), say for a bundle $B$ of type $q$,
note that $s_B(\o x) = t'_{pq}(s_{B_1}(\o x),\ldots,s_{B_{k_\er}}(\o x))$,
where $B_1,\ldots,B_{k_\er}$ form the $k_\er$ element group of bundles (of type $p$)
that were passed through $\Theta'_{pq}$.
Since $t'_{pq}$ returns boolean values on boolean arguments, we get (iv).

To prove (v) let $C_1,\ldots,C_{\er-1}$ be the bundles
resulting from passing the $k_\er$-element group
$B_1,\ldots,B_{k_\er}$ of bundles of type $p$
through $\er-1$ bunches $\Theta'_{pq}$ (with $q\neq p$).
If $C_s$ is of type $q$ then
$s_{C_s}(\o x) =t'_{pq}(s_{B_1}(\o x),\ldots,s_{B_{k_\er}}(\o x))$
and consequently
\[
\pand(s_{C_1}(\o x),\ldots,s_{C_{\er-1}}(\o x)) =
\pand\set{t'_{pq}(s_{B_1}(\o x),\ldots,s_{B_{k_\er}}(\o x)) : q\neq p}.
\]
Due to the equation (\ref{eqn:and}) the last conjunction is equal to
$\pand(s_{B_1}(\o x),\ldots,s_{B_{k_\er}}(\o x))$.
Thus the two conjunctions of all the sums of the form $s_B(\o x)$:
one before processing the bundles through a given level
and the other one after processing them, are equal.
This shows (v).

After arriving at the level $h-2$, our circuit $\nabla_{h-2}$ produces
$b_{h-2}= \er\err\cdot k^{\es-1}=\er k_\es$ bundles, i.e. $k_\es$ bundles in each type.
Now we put all these $k_\es$ bundles of one type, say $p$,
into one group and proceed this group through $\es-1$ bunches $\Theta'_{pq}$
with $q$ ranging over some $\es-1$ element subset $Q_p\ci \set {q\neq p : q\geq \er}$.
Again, as in the proof of invariant (v), we argue that
$\pand(s_{C_1}(\o x),\ldots,s_{C_{\es-1}}(\o x)) =
\pand(s_{B_1}(\o x),\ldots,s_{B_{k_\es}}(\o x))$
where $C_1,\ldots,C_{\es-1}$ are the bundles
resulting from passing $k_\es$-element group $B_1,\ldots,B_{k_\es}$
of bundles of type $p$ through the bunches $\Theta'_{pq}$ with $q\in Q_p$.
The output of the $(h-1)$-th level consists of $\er\ess$ bundles,
as each of the $\er$ groups is passed through $\es-1$ bunches $\Theta'_{pq}$
with large primes $q$.
On the other hand for a fixed large $q$ at most $\er-1$ primes $p\neq q$
may contribute to the bunches $\Theta'_{pq}$ that are actually used on level $h-1$.
To distinguish those primes we put
$\zz_j=\set{i : \Theta'_{p_i p_j} \mbox{\ is used on level \ } h-1}$ for $j\leq \es$.
Note that
$\set{1,\ldots,\es}-\set{j} \ci \zz_j \ci \set{1,\ldots,\er}-\set{j}$,
i.e. in particular $\card{\zz_j} \leq \er-1 < p_j$.
In this notation we enumerate all the $\card{\zz_{1}}+\ldots+\card{\zz_\es}$ bundles resulting from level $h-1$ by
\(
C_{1}^1,\ldots,C_{1}^{\card{\zz_{1}}},
C_{2}^1,\ldots,C_{2}^{\card{\zz_{2}}},\ldots\ldots,
C_{\es}^1,\ldots,C_{\es}^{\card{\zz_{\es}}}.
\)
Denoting $s_{C^i_j}(\o x)$ simply by $s^i_j(\o x)$ we now express the property (v) as
\begin{equation}
\label{eqn:s-and}
\pand(x_1,\ldots,x_n)=\pand\set{s^i_j(\o x) : j\leq\es \mbox{ and } i\in\zz_j}.
\end{equation}
Now, at the very last level we put all $\card{\zz_{1}}+\ldots+\card{\zz_\es}$ bundles,
with $C^i_j$ being repeated $\frac{m}{p_j}$ times,
into the gate $\cmod_{m}^\set{\zzz}$,
where $\zzz =\sum_{j\leq\es} \frac{m}{p_j}\cdot\card{\zz_j} \mod{m}$.
This gate computes (modulo $m$) the sum
\[
S(\o x) = \sum_{j\leq\es} \sum_{i\in \zz_j} \frac{m}{p_j}\cdot s^i_j(\o x)
\]
and turns it to $1$ if $S(\o x)=\zzz$ and to $0$ otherwise.
Thus, due to (\ref{eqn:s-and}),
we are left with showing that $S(\o x)=\zzz$
iff $s^i_j(\o x)=1$ for all $j\leq\es$ and $i\in \zz_j$.
Obviously if all the $s^i_j(\o x)$'s are $1$ then the sum $S(\o x)$ is $\zzz$.
Conversely, as in the proof of Proposition \ref{prp:force-narrow}, we first identify the direct sum
$\bigoplus_{j=1}^\es \frac{m}{p_j} \cdot \mathbb{Z}_{p_j}$ with a subgroup of $\mathbb{Z}_m$.
Then the assumption that
$\zzz= S(\o x) = \sum_{j\leq\es} \frac{m}{p_j}\cdot\sum_{i\in \zz_j} s^i_j(\o x)$
together with the fact that
$0\leq \sum_{i\in \zz_j} s^i_j(\o x) \leq \card{\zz_j} \leq \er-1 <p_j$
gives, by the unique decomposition in the direct sum,
that for each $j\leq\es$ we have $\sum_{i\in \zz_j} s^i_j(\o x)=\card{\zz_j} \mod p_j$.
But now $s^i_j(\o x)\in\set{0,1}$ and $\card{\zz_j}<p_j$ yield that all the $s^i_j(\o x)$'s are $1$.

\medskip

It remains to calculate the size of the entire circuit.
Each of the first $h-2$ levels has $2^{O(p_i^{\niu_i} \log k_\er)}$ gates
in each bunch of type $p_i$.
There are at most $O(n)$ bunches of each type.
Using $p_i^{\niu_i} \leq p_i k_\er^{1/\err}\in O(k)$
we bound the size of each bunch by $2^{O(k\log k)}$.
The same holds on the level $h-1$.
Summing up we bound the size of entire circuit by
$2^{O(k\log n)} \leq 2^{O(n^{1/(\err(h-2)+\ess)} \log n)}$.
\end{proof}

\bigskip

Now we are ready to show Theorem \ref{thm:omegabar}
that, in comparison to Proposition \ref{prp:omega-omegabar},
increases the degree of the root just by one.

\begin{proof}
Our circuits here are based on those from the proof of Proposition \ref{prp:omega-omegabar}
by modifying only two levels: $\nabla_0$ and $\nabla_1$.
This time we start with assuming that $n=k^{\err(h-2)+\es}$ for some integer $k$.
Also, additionally to the $\niu_j$'s and the $\o\niu_j$'s
(exactly as in the proof of Proposition \ref{prp:omega-omegabar})
we pick $\niu^0_j$ to satisfy $p_j^{\niu^0_j-1} \leq k < p_j^{\niu^0_j}$ for all the $j$'s.
The starting circuit $\nabla_0$ takes $n$ inputs $x_1,\ldots,x_n$
and makes $b_0= \er\cdot n$ bundles,
each of which consisting of one typed variable, so that there are exactly $n$ bundles in each type.
To proceed these bundles through the gates of $\nabla_1$ we will group $n$ bundles in each type
into groups of size $k_0=k^\er$, but in a synchronized way.
By this synchronization we mean that first the set $\set{1,\ldots,n}$ is split into $n/k_0$ groups
$G_i$ of size $k_0$
and then in each type, say $p$, we form a group of bundles $G^p_i = \set{x_j : j\in G_i}$.
Next, each such group $G^p_i$ is passed though all the $\Theta'_{pq}$'s (with $q\neq p$)
to get the bundles $\Theta'_{pq}(G^p_i)$.
As previously we want to have that $\pand(x_1,\ldots,x_n)$ coincides with the conjunction of all the
$s_B(\o x)$'s with $B$ ranging over all the bundles produced by $\nabla_1$, i.e. that:
\[
\pand(x_1,\ldots,x_n)=\pand\set{s_{\Theta_{pq}(G^p_i)}(\o x) : \ p\neq q, \ i=1,\ldots,n/k_0}.
\]
We get this by observing that $\pand(G^p_i)$ can be replaced by the conjunction of $s_B(\o x)$ for (at least $\er$) bundles $B$ of all $\er$ different types $p_1,\ldots,p_\er$.
This however is witnessed  by
\[
\pand(G^p_i)=\pand\left(\set{s_{\Theta_{pq}(G^p_i)}(\o x) : \ q\neq p}\cup \set{s_{\Theta_{qp}(G^q_i)}(\o x) : \ q\neq p}\right),
\]
due to the fact that for a fixed $i$ our synchronization spans the sets $G^p_i$ and $G^q_i$ on the very same variables.

In this process $\nabla_1$ replaces each group of $k_0=k^\er$ bundles
by $\er-1$ new bundles.
This means that $\nabla_1$ produces
$b_1= \err\cdot\frac{b_0}{k^\er}=\err\er\cdot k^{\err(h-3)+\ess}$ bundles,
which is exactly the number of bundles produced by $\nabla_1$ in the proof of Proposition \ref{prp:omega-omegabar}.
This allows us to put these bundles into the consecutive levels of the circuit described in that proof.

As previously our choice of 
$k_0, k_\er, k_\es$
(for determining the sizes of the groups of bundles)
yields that, on each level, the sizes of the bunches used in our circuit
are bounded by $2^{O(k\log k)}$.
Combining this with the fact that on each level at most $O(n)$ bunches are used
and with $n=k^{\err(h-2)+\es}$ we get that our circuit has the size
bounded by $2^{O({n}^{1/(\err(h-2)+\es)}\log n)}$.
\end{proof}

\section{Probabilistic circuits}
\label{sec:cc2-prob}

In this section we prove Theorem \ref{thm:random-and},
i.e. we construct polynomial size $\cc{p;q}$-circuits $\Gamma_n$ computing $\pand_n$
with the help of $\rbit=6+\log n$ additional random bits.
This means that $\Gamma_n$ has $n+\rbit$ inputs and
for each $n$-tuple $\o a \in \set{0,1}^n$
for at least $\frac{2}{3}$ possible tuples $\o b \in \set{0,1}^\rbit$
we have $\Gamma_n(\o a, \o b) = \pand_n(\o a)$.
These circuits will be based on $O(\rbit)$-ary special $\zpq$-expressions
so that we can control their size to be polynomial in $n$, i.e. $2^{O(\rbit)}$.

To start our construction define $\Lambda$ to be the set of all tuples
$\lambda=\left(\lambda_{\o c,j}\right)^{\o c \in \set{0,1}^\rbit}_{j=1,\ldots,\kap\rbit}$
of length $2^\rbit \kap\rbit$, where $\kap =\lceil\log_{\frac{p}{p-1}}2\rceil$
and each $\lambda_{\o c,j}$ is an $GF(p)$-affine combination of the $x_i$'s
satisfying $\lambda_{\o c,j}(1,\ldots,1)=1$.
Define $\bb'(z)=1-\bb(z)$ so that for $\lambda \in \Lambda$ we put
\[
t_\lambda(\o x,\o b) =
\sum_{\o c \in\set{0,1}^\rbit}
\prod_{i=1}^{\rbit} \bb'(b_i-c_i) \cdot
\prod_{j=1}^{\kap\rbit} \bb(\lambda_{\o c,j}(\o x)),
\]
to show that
\begin{itemize}
  \item each $t_\lambda(\o x,\o b)$ can be turned into $\zpq$-expression with $2^{O(\rbit)}$ summands (corresponding to the number of gates in the circuits realizing this expression),
  \item for at least one $\lambda\in \Lambda$ the expression $t_\lambda(\o x,\o b)$ calculates $\pand_n(\o x)$ for at least $\frac{2}{3}$ of the $\o b$'s in $\set{0,1}^\rbit$.
\end{itemize}
For the first item note that each summand in $t_\lambda(\o x,\o b)$
can be obtained by an appropriate substitution in a $(\rbit+\kap\rbit)$-ary function
\(
Z_p^\rbit \times Z_p^{\kap\rbit} \ni (\o u, \o z) \mapsto
\bb'(u_1)\cdot\ldots\cdot\bb'(u_\rbit)\cdot\bb(z_1)\cdot\ldots\cdot\bb(z_{\kap\rbit}) \in Z_q.
\)
By Fact \ref{trzy-jeden} such function can be represented by a $\zpq$-expression with
$O(p^{(\kap+1)\rbit})$ summands.
Now, summing up (modulo $q$) over the $\o c$'s we end up with a $\zpq$-expression with
$O(2^\rbit p^{(\kap+1)\rbit})=2^{O(\rbit)}= \poly(n)$ summands.

Before showing the second item note that for fixed $\o b \in \set{0,1}^\rbit$ the expression
$t_\lambda(\o x,\o b)$ reduces to only one summand, namely
$\prod_{j=1}^{\kap\rbit} \bb(\lambda_{\o b,j}(\o x))$.
Now, for a fixed $\o a \in \set{0,1}^n$ and $\o b \in \set{0,1}^\rbit$
the random variable $X_{\o a,\o b}$  checks for a particular tuple
$(\lambda_{\o b,j})_{j=1,\ldots,\kap\rbit}$
if the value $\prod_{j=1}^{\kap\rbit} \bb(\lambda_{\o b,j}(\o a))$ coincides with $\pand_n(\o a)$.
Thus the sum $X_{\o a} = \sum_{\o b \in\set{0,1}^\rbit} X_{\o a,\o b}$,
defined now on entire $\Lambda$,
simply counts the number of the $\o b$'s for which $t_\lambda(\o a,\o b) = \pand_n(\o a)$.
We conclude our argument with showing that
$\pr{\mmeet_{\o a \in \set{0,1}^n} X_{\o a}\geq \frac{2}{3}\cdot 2^\rbit}\neq 0$.
Note that for fixed $\o a \neq \o 1$ and randomly chosen $\lambda_{\o c,j}$
we have $\pr{\lambda_{\o c,j}(\o a)\neq 0} =\frac{p-1}{p}$ so that
$\pr{X_{\o a,\o b}=0}= \left(\frac{p-1}{p}\right)^{\kap\rbit}=2^{-\rbit}$ and
$E(X_{\o a,\o b})=1-2^{-\rbit}$.
Consequently $E(X_{\o a})=2^\rbit(1-2^{-\rbit})=2^\rbit -1$.
Fixing $\delta$ so that $(1-\delta)E(X_{\o a})=\frac{2}{3}\cdot 2^\rbit$
we apply Chernoff's inequality for the lower tail to get
$\pr{X_{\o a}\leq \frac{2}{3}\cdot 2^\rbit}
\leq \exp\left(-\frac{E(X_{\o a})\cdot\delta^2}{2}\right)
\leq \exp\left(-\frac{64n-1}{32}\right)
<2^{-n}$.
Consequently probability of the fact that no $\lambda \in \Lambda$ leads to $t_\lambda$ with desired property is bounded by
$\pr{\jjoin_{\o a \in \set{0,1}^n} X_{\o a}\leq \frac{2}{3}2^\rbit} <2^n\cdot2^{-n}=1$,
as required.

\section{Algorithms}
\label{sec:algo}

In Sections \ref{sec:cc2} and \ref{sec:cchm} we have seen how to construct subexponential conjunctions and how it helps to encode 3-CNF SAT in satisfiability of modular circuits.
Obviously better upper bounds for the size of circuits realizing \pand \
(and consequently 3-CNF formulas) give rise to higher complexity of $\cch{h}{m}$-SAT.
In particular a polynomial upper bound for the size of \pand would show \npc{ness} of $\cch{h}{m}$-SAT.
Although, in Section \ref{sec:cc2-prob} we have shown that \pand can be realized by a
probabilistic $\cch{h}{m}$-circuits of polynomial size (provided $h,\omega(m)\geq 2$),
we strongly believe that this cannot be done without those random bits.

In this section we analyze how the lower (superpolynomial) bound for the size of circuits realizing \pand \ can be used to (subexponentially) bound the complexity of $\cch{h}{m}$-SAT from above.

To this end for fixed depth $h$ and modulus $m$ by $\gamon(n)$
we denote the size of the smallest possible $\cch{h}{m}$-circuit computing $\pand_n$.
Note first that (according to Proposition \ref{shallow-narrow-and})
if $h=1$ or $\omega(m)=1$
the values $\gamon(n)$ are defined only for finitely many first integers $n$.
However, independently of $h$ and $m$,
Fact \ref{trzy-jeden} ensures us that $\gamon$ is at most exponentially large
and therefore computable in \expexp. In our considerations we need much better bound for the time needed to compute $\gamon(n)$.
Note that the functions bounding sizes of the circuit constructed in Propositions
\ref{prp:force-narrow}(\ref{force-narrow-and}),
\ref{prp:single-primes},
\ref{prp:r-and-h-pol},
\ref{prp:omega-omegabar}
and Theorem \ref{thm:omegabar}
are of the form $2^{O(n^{1/\delta}\log n)}$ and can be computed in \ptime.
Although we cannot guarantee that $\gamon$ is \ptime-computable,
it would be enough for us to bound it 
from below by such a function
(which is still close enough to $\gamon$).

Now we provide two algorithms for satisfiability of $\cch{h}{m}$-circuits,
a deterministic one and a slightly faster randomized one
with running times depending on the growth rate of $\gamon$,
or rather it inverse.
For a partial increasing function $f: \mathbb{N} \map \mathbb{N}$ by $\pre f k$ we mean the largest $n$ with $f(n)\leq k$.

\begin{thm}
\label{thm:algo}
Suppose that $\gamon$ has \ptime-computable increasing lower bound $f$.
Then there are two algorithms for checking if an $n$-ary $\cch{h}{m}$-circuit is satisfiable:
\begin{itemize}
  \item a deterministic one with the running time
    $O\left(\poly\card{\Gamma} +  2^{\pree f {\card{\Gamma}}\cdot\log n} \cdot \card{\Gamma}\right)$,
  \item a randomized one with the running time
    $O\left(\poly\card{\Gamma} +  2^{\pree f {\card{\Gamma}}} \cdot \card{\Gamma}\right)$.
\end{itemize}
\end{thm}

\begin{proof}
Our deterministic algorithm is based on a brute-force search for a satisfying tuple in a relatively small set $S$ of size $n^{\pree f {\card{\Gamma}}}$
consisting of all the tuples $a\in \set{0,1}^n$ with at most $\pree f {\card{\Gamma}}$ ones.
To determine this set we first need to know the value $\pree f {\card{\Gamma}}$.
But since $f$ is \ptime-computable this can be done in $\poly\card{\Gamma}$ steps.
This together with checking whether $S$ contains a satisfying tuple takes
$\poly\card{\Gamma} + O\left(\card{\Gamma} \cdot n^{\pree f {\card{\Gamma}}} \right)$ steps, as claimed.

\medskip
Since $\pree \gamon {} \leq \pree f {}$ we are left with showing that if $\Gamma$ is satisfiable
then it can be satisfied by a tuple $a\in \set{0,1}^n$
with $\card{\pre a 1} \leq \pree \gamon {\card{\Gamma}}$.
Suppose then that $a$ is a non-zero satisfying tuple with the minimal number of ones.
By this minimality we know that $\Gamma$ with all the inputs with indices outside $\pre a 1$
set to $0$ behaves like the $\card{\pre a 1}$-ary \pand.
Thus $\gamon \card{\pre a 1} \leq \card{\Gamma}$
so that the tuple $a$ has at most $\pree \gamon {\card{\Gamma}}$ ones.

\medskip

On the other hand our second algorithm, the probabilistic one, is based on randomly choosing sufficiently many inputs so that the probability of having a satisfying one among them
exceeds $1/2$, if there is any such satisfying tuple at all.
We claim that $2^{\pree \gamon {\card{\Gamma}}}$ samples suffices.
Indeed, if $\Gamma$ is constant then any single sample witnesses its (un)satisfiability.
Remark \ref{rmk-reducing-stack}  allows us to modify a nonconstant circuit $\Gamma$
to get $(n-k)$-ary spike circuit $\Gamma'$ for some $k\leq \log\card{\stack{\Gamma}}$,
so that $\card{\Gamma'}\geq \gamon(n-k)$.
Consequently $\card{\Gamma}\geq \card{\Gamma'}\geq \gamon(n-\log\card{\stack{\Gamma}})$,
which together with $\card{\pre \Gamma 1} \geq \card{\stack{\Gamma}}$ gives
${\card{\pre \Gamma 1}}/{2^n} \geq 2^{-\pree \gamon {\card{\Gamma}}}$.
This simply means that 
we will find a tuple from $\pre \Gamma 1$
among $2^{\pree \gamon {\card{\Gamma}}}$ samples.
But again, to calculate how long we need to sample we increase $2^{\pree \gamon {\card{\Gamma}}}$
to $2^{\pree f {\card{\Gamma}}}$
and use the fact that $f$ is \ptime-computable.
\end{proof}

From our proof of Theorem \ref{thm:algo} we get the following generalization of Corollary \ref{cor:bal1}.

\begin{cor}
\label{cor:bal2}
The balance of a $\cch{h}{m}$-circuit $\Gamma$ is at least $2^{1-\pree \gamon {\card{\Gamma}}}$.
\end{cor}

Observe here, that like in Corollary \ref{cor:lan-bal1},
we can use the function $\gamon$ to bound from below the number of words
(of a given length) in a language recognizable by polynomial size $\cch{h}{m}$-circuits.
In particular the suspected lower bound for $\gamon$ of the form $2^{\Omega(n^\delta)}$
translates into the bound $2^{n-O(\log^{1/\delta} n)}$.

Although our random sampling algorithm RanSam described in the proof of Theorem \ref{thm:algo}
is not involved, the proof itself tells that a bigger lower bound for $\gamon$ allows us to reduce the number of samples in RanSam. Below we show that this connection is two-sided.

\begin{prp}
\label{prp:ransam}
If RanSam works (with probability at least $1/2$) with at most $2^{f\card{\Gamma}}$ samples
for some increasing computable function $f$,
then $\pre f n \leq \gamon(n+1)$.
\end{prp}

\begin{proof}
We run RanSam on the circuit $\pand_n$ with $2^{f\gamon(n)}$  samples,
so that the expected number of satisfying tuples is $2^{f\gamon(n)}/2^n$.
This procedure however has to find, with probability at least $1/2$
the unique satisfying tuple.
Thus, Markov inequality yields $2^{f\gamon(n)}/2^n \geq 1/2$
so that $\pre f {n-1} \leq \gamon(n)$.
\end{proof}

Combining Theorem \ref{thm:algo} and Proposition \ref{prp:ransam}
we get that the suspected lower bound $2^{\Omega(n^\delta)} \leq \gamon$
is equivalent to the upper bound $2^{O(\log^{1/\delta}\card{\Gamma})}$ for the running time of RanSam.
Actually any superpolynomial lower bound for $\gamon$ mutually translates into
substantially subexponential (i.e. at most $2^{\card{\Gamma}^{o(1)}}$) number of samples.

As for now only slightly superlinear lower bounds $\Omega(n\cdot \epsi(n))$ for $\gamon$ are known,
as $\epsi(n)$ is an extremely slowly increasing function
(see \cite{LowBounds}).
Although the functions $\epsi(n)$ depend on $h$ and $m$,
a careful inspection of their description 
shows that the inverse to $n\cdot \epsi(n)$ is always bounded by
$O(n/\epsi(n))$.
This, together with Theorem \ref{thm:algo} shows the following result.

\begin{thm}
\label{thm:cc-vs-ac}
Satisfiability of $\cch{h}{m}$-circuits $\Gamma$
is solvable in probabilistic $2^{O(\card{\Gamma}/\epsi(\card{\Gamma})}$ time.
\end{thm}

This Theorem stays in a big contrast to the lower bound $2^{\Omega(\card{\Gamma})}$
(provided by the randomized version of ETH \cite{rETH09}) for
probabilistic algorithms for satisfiability of $AC^0$-circuits.

\medskip
We conclude this section with arguing that
(under some additional assumption about effective coding of 3-CNF formulas by modular circuits)
the running time of RanSam is hard to beat.
Our heuristic assumption simply says that there is a \ptime algorithm that turns 3-CNF formulas $\Phi$
with $\cll$ clauses
into $\cch{h}{m}$-circuits of size bounded by $O(\gamon(c\cll))$.
Assuming also that $\gamon$ (or some of its $\Theta$-equivalents) is \ptime-computable
we know that RamSam runs with $2^{O(\pree \gamon {\card{\Gamma}})}$ samples.
On the other hand \ethhh,
applied to the circuit $\Gamma$ produced from 3-CNF formula
by the algorithm supplied by our heuristic assumption,
gives an integer $d>0$ so that $\cch{h}{m}$-SAT
cannot be solved in $O(2^{\frac{1}{d}\pree \gamon {\card{\Gamma}}})$.
Thus the best imaginable algorithm solving $\cch{h}{m}$-SAT
has the running time bounded by 
a polynomial applied to the running time of RanSam.

\section{Concluding remarks and applications}

In view of the results in Section \ref{sec:cchm},
in particular a spectacular role played by $\varpi(m)$,
as well as the easiness of increasing the degree of the root just by $1$,
it seems to be really hard to state reasonable conjectures
for the asymptotic behaviour of the $\gamon$'s.
As for now, for $h=2$ the degree of the root (occurring in the exponent)
is at least $\omega(m)$ (Proposition \ref{prp:force-narrow})
and for $h\geq 3$ at least $\err(h-2)+\es$ (Theorem \ref{thm:omegabar}).
However for the `majority' of potential moduli $m$
we know that $\varpi(m)$ is pretty close to $\omega(m)$,
so that this degree is almost $\err(h-1)+1$
(and coincides with $\omega(m)$, whenever $h=2$).
Due to the fact that prime factorization (i.e. the number $\omega(m)$)
may contribute fully into this degree
and the depth $h$ contributes by the factor $h-1$,
it seems natural to suspect that the bound for $\log\gamon(n)$ could be of the form
$n^{1/(\er(h-1))}\log n$.

Another remark we want to make here is the difference between the circuits of the form
$\cch{h}{p;q;p;q;\ldots}$ (with $p\neq q$) and $\cch{h}{p\cdot q}$.
In the later case we have $\es=\er=2$ so that the bound for the considered degree is $h$,
while Proposition \ref{prp:single-primes} gives degree $h-1$ in the first case.
Moreover, it seems that there is no room for improving this $h-1$ in this case.

This difference in locations of primes on different levels is even more striking
for $\cch{2}{p;m}$ and $\cch{2}{m;p}$,
whenever $m$ has $r\geq 2$ prime divisors except $p$.
In the first case we can actually argue,
as in the proof of Proposition \ref{prp:force-narrow},
to get the upper bound $2^{n^{1/r}\log n}$ for $\gamon$,
while \cite{BST90,ST06} give $2^{\Omega(n)}$ lower bound in case of $\cch{2}{m;p}$.

\medskip
The technique we have developed for proving Proposition \ref{prp:force-narrow}
can be used to determine (modulo \ethhh) the complexity of solving equations over the dihedral groups $\m D_{2k+1}$,
i.e. groups of symmetries of regular polygons with odd number of sides.
Some of the variables in these equations are already preevaluated
(as otherwise every equation has a trivial solution with all the variables set to the neutral element of the group).
This is equivalent to consider polynomials (instead of terms) over groups.
The decision version of this problem for the group $\m G$ is denoted by $\polsat{G}$.
Analogously by $\poleqv{G}$ we mean the problem of deciding whether two polynomials over $\m G$ define the same function.
Note here that from the paper \cite{goldman-russell} of Goldmann and Russell
we know that \polsat{} is \npc for nonsolvable groups and in \ptime for nilpotent groups.
Moreover the paper \cite{ikkw} partially fills this gap
by showing that (modulo \ethhh) $\polsat{G}$ is not in \ptime unless
$\m G$ has Fitting length at most $2$,
i.e. $\m G$ is a wreath product of two nilpotent groups.
This paper refutes a long standing belief that \polsat{} for all solvable groups is in \ptime.
The conjecture was based on many examples of groups that are in fact 2-nilpotent.
The very recent paper of Földvári and Horváth \cite{foldvari-horvath}
summarizes most of these examples by showing that $\polsat{\m G}$ is in \ptime
whenever $\m G$ is a semidirect product of a $p$-group and an abelian group.
Note here that the dihedral groups $\m D_{p^k}$, with prime $p$, fall into this realm.
On the other hand our characterization below dismisses such a speculation
about tractability of \polsat{} for groups of Fitting length 2 (unless \ethhh fails).

\begin{thm}
\label{thm:dm}
If \ethhh holds then for each odd integer $m\geq 3$ the problem $\polsat{D_m}$ is in \ptime
iff $\omega(m)=1$.
\end{thm}

\begin{proof}
Remind that the dihedral group $\m D_m$ is generated by two elements,
a rotation $\rho$ (with angle $2\pi/m$)
and a reflection  $\sigma$ satisfying
$\rho^m=1, \sigma^2=1$ and $\sigma\rho=\rho^{-1}\sigma$.
This means that $\m D_m$ has $2m$ elements:
$m$ rotations $\rho^0, \rho^1, \rho^2,\ldots, \rho^{m-1}$
and $m$ reflections $\sigma, \sigma\rho, \sigma\rho^2,\ldots, \sigma\rho^{m-1}$.

If $\omega(m)=1$ 
then we have already noted
that  \cite{foldvari-horvath} puts $\polsat{D_{m}}$ into \ptime.

Now suppose $m=p_1^{\alf_1}\cdot\ldots p_\er^{\alf_\er}$,
where the $p_j$'s are pairwise different  odd primes and $\er\geq 2$.
Since  the rotations form a cyclic group isomorphic to $\mathbb{Z}_m$
for each $j=1,\ldots,\er$ there is a rotation, say $\rho_j$,
generating a cyclic subgroup of order $p_j$.

Define unary polynomials (with $j=1,\ldots,\er$) by putting
$\ee(x) = \sigma(\sigma x^m)^m, \ee_j(x) = x^{2m/p_j}$
and $\bb_j(x) = (\rho_j \ee(x) \rho_j^{-1} \ee(x)^{-1})^{\frac{m+1}{2}}$
and observe that the range of $\ee$ is $\set{1,\sigma}$,
i.e. the group isomorphic to $\mathbb{Z}_2$,
while $\ee_j$ maps the group $\m D_m$ onto its cyclic subgroup
$\set{1,\rho_j,\rho_j^2,\ldots,\rho_j^{p_j-1}}$ isomorphic to $\mathbb{Z}_{p_j}$.
Moreover the polynomial $\bb_j$ maps the group $\set{1,\sigma}$ onto
$\set{1,\rho_j} \ci \set{1,\rho_j,\rho_j^2,\ldots,\rho_j^{p_j-1}}$
and therefore $\bb_j$ can be used to build $\mathbb{Z}[2,p_j]$-expressions
as polynomials of $\m D_m$.

Now we adapt the proof of Proposition \ref{prp:force-narrow}(1) to our setting.
For a 3-CNF formula $\Phi$ we borrow the $\mathbb{Z}[2,p_j]$-expressions by putting
$t^\Phi_j=t^\Phi_{2,p_j}$ to build the polynomial $T^\Phi$,
but we modify the original definition (\ref{tfi}) to be read
\[
T^\Phi(x_1,\ldots,x_n) = \sum_{j=1}^{\er} \ t^\Phi_j(x_1,\ldots,x_n)
\]
where the sum is computed in the direct sum $\bigoplus_{j=1}^\er \mathbb{Z}_{p_j}$
identified with a subgroup of the group $\mathbb{Z}_m$ of all rotations.
Now we simply transform 3-CNF formula $\Phi$ into the equation $T^\Phi(\o x)=0$,
with $0$ being the neutral element of both $\mathbb{Z}_m$ and $\m D_m$.
To see that $\Phi$ is satisfiable iff the corresponding equation has a solution in $\m D_m$,
we simply go back and forth between the boolean values and the elements of $\m D_m$
by identifying
the rotations, i.e. elements of
$\ee_0^{-1}(1)$ with the boolean value true
and the reflections, i.e. elements  of $\ee_0^{-1}(\sigma)$ with the boolean value false.

Obviously, as previously, the length of $T^\Phi$ is bounded by $2^{O(\sqrt[\er]{\cll}\log \cll)}$
where $\cll$ is the number of clauses in $\Phi$.
Thus \ethhh yields that $\polsat{D_m}$ cannot be in \ptime.
\end{proof}

An analysis of the complexity for \polsat{} over all dihedral groups $\m D_m$
is postponed to our paper \cite{ikk:dihedral}.
In particular our method used in Theorem \ref{thm:dm} is applied to a more subtle situation where
$m$ is even but has at least two different odd prime divisors.

Another feature of the dihedral groups $\m D_m$ is that $\poleqv{D_m}$ is in \ptime
for all $m$, see \cite{burris-lawrence}.
Thus Theorem \ref{thm:dm} provides the first examples of finite groups with tractable \poleqv{}
and untractable \polsat{} (modulo \ethhh).
Note here  that every group with tractable \polsat{} has tractable \poleqv{},
as to decide whether two polynomials $t,s$ are equal
we simply check that none of the $\card{G}-1$ equations of the form $ts^{-1}=a$
(with $a$ ranging over $G-\set{1}$) has a solution.

\medskip

Almost the same argument can be used in the setting of multivalued circuit
satisfiability \csat{} and circuit equivalence \ceqv{}, as defined in  \cite{ik:lics18}.
Such multivalued circuits are built over a fixed finite algebra $\m A$
so that the gates here simply compute the basic operations of the algebra.
The paper \cite{ik:lics18}
initiated
a systematic project of characterizing finite algebras $\m a$ with
$\csat{\m A}$ in \ptime and provided a partial characterization for algebras from congruence modular varieties.
However a somehow similar (to \polsat{} for groups) gap was left open,
namely the unsolved complexity of \csat{} and \ceqv{} for nilpotent but not supernilpotent algebras.
In paper \cite{ikk:lics20} we constructed algebras $\dpp{p_1,\ldots,p_h}$ built over the alternating chain of primes $p_1 \neq p_2 \neq p_3 \neq \ldots \neq p_h$
with \csat{} and \ceqv{} outside \ptime,
provided $h\geq 3$ and \ethhh holds.
Later the paper \cite{komp:mfcs21} developed 
these methods to actually force nilpotent algebras with \csat{} or \ceqv{} in \ptime to be wreath products of two supernilpotent algebras.
On the other hand \cite{ikk:mfcs18} provides examples of such wreath products (that are actually 2-nilpotent) with \csat{} and \ceqv{} in \ptime.
Although \ceqv{} for all 2-nilpotent algebras has been confirmed \cite{kkk} to be in \ptime,
an analogue for \csat{} is blocked by the following example,
the proof of which simply repeats the argument for Theorem \ref{thm:dm}.

\begin{ex}
\label{ex:dqqq}
\csat{} for the following $2$-nilpotent algebras is outside \ptime, modulo \ethhh:
\\
For any sequence $p_0,p_1,p_2,\ldots,p_\er$ of pairwise different primes
the algebra $\dhr$ is the group
$\m Z_{p_0}\times \m Z_{p_1} \times \ldots \times \m Z_{p_\er}$
endowed with $2\er+1$ unary operations
\[
\begin{array}{ccll}
\ee_j(x_0,x_1,\ldots,x_\er) &=& (0,\ldots,0,x_j,0,\ldots,0), &\mbox{ for $j=0,1,\ldots,\er$,}\\
\bb_j(x_0,x_1,\ldots,x_\er) &=& (0,\ldots,0,b^*_j(x_{0}),0,\ldots,0), &\mbox{ for $j=1,\ldots,\er$,}
\end{array}
\]
where $b^*_j : \m Z_{p_0} \map \m Z_{p_{j}}$ is the function given by
$b^*_j(0)=0$ and $b^*_j(a)=1$ otherwise.
\hfill\myqed
\end{ex}

\section{Easy stuff}
\label{sec:easy}

\subsubsection*{Proof of Proposition \ref{shallow-narrow-and}}

To warm up, note that for any modulus $m$ and $n\geq m$
each sequence $\alpha_1,\ldots,\alpha_n$ of integers contains a nonempty subsequence
$\alpha_{i_1},\ldots,\alpha_{i_k}$ that modulo $m$ sums up to $0$.
Indeed, either there is $0$ among the sums
$\alpha_1,
\alpha_1+\alpha_2,
\ldots,
\alpha_1+\ldots+\alpha_n$
or at least two of them are equal,
making their difference , i.e. a shorter nonempty sum, to be $0$.

Now, the only (say $n$-ary) gate $\cmod_m^{A}$ in the $\cch{1}{m}$-circuit (that takes $\alpha_i$ times the input $x_i$),
after checking if $\sum_{i=1}^n \alpha_i x_i$ belongs to $A$,
returns the very same value on the constant sequence $x_1=\ldots=x_n=1$
and its modification obtained by switching $x_{i_1},\ldots,x_{i_k}$ to $0$.
This destroys the possibility for $\cmod_m^{A}$ with $n \geq m$ inputs to serve as $\pand_n$.

\medskip
For $\cch{h}{p^k}$-circuits  we induct on $h$ to show how from a particular circuit $\Gamma$
pass to a polynomial $w_\Gamma(\o x)$ over $GF(p)$ so that this polynomial:
\begin{itemize}
  \item computes the circuit $\Gamma$,
  \item is presented in its sparse representation,
  \item contains monomials of degree $d^h$ for some constant $d$ depending on $p^k$ only.
\end{itemize}
Having done that, we pick a monomial in $w_\Gamma(\o x)$ of minimal degree and evaluating all the $x_i$'s occurring in this monomial by $1$ and the other $x_i$'s by $0$,
we know that $w_\Gamma(\o x)\neq 0$.
However $n>d^h$ ensures us that at least one of the $x_i$'s is $0$,
contrary to the fact that $\Gamma$ is supposed to compute $\pand_n$.

To start our induction for $h=1$ we refer to the paper \cite{bt} of Beigel and Tarui,
where Lemma 2.1 supplies us with a polynomial $r_0(x_1,\ldots,x_n)$ over $GF(p)$
that on the boolean values of the $x_i$'s behaves as the gate $\cmod_{p^k}^{\set{0}}$.
In fact in the proof of that Lemma it is shown that the polynomial
\[
r_0(x_1,\ldots,x_n) =
\prod_{j=1}^{k-1}
\left(
1 - \left(
\sum_{I\ci\set{1,\ldots,n}, \card{I}=p^j} \quad \prod_{i\in I} x_i
\right)^{p-1}
\right)
\]
does the job.
It is easy to see that the degree $d_0$ of $r_0(x_1,\ldots,x_n)$ is bounded by $p^{k+1}$, independently of $n$.
Since on the boolean values of the $x_i$'s the polynomial $r_0(\o x)$ can be represented as
\[
r_0(x_1,\ldots,x_n) =
\prod_{j=1}^{k-1}
\left(
1 - \binom{\sum_{i=1}^n x_i}{p^j}^{p-1}
\right)
\]
we get that $r_c(x_1,\ldots,x_n)=r_0(x_1-c,x_2,\ldots,x_n)$ computes the gate $\cmod_{p^k}^{\set{c}}$.
Consequently $r_A(\o x) = 1 - \prod_{c\in A} \left(1-r_c(\o x)\right)$ computes $\cmod_{p^k}^{A}$.
The degree of the polynomial $r(\o x)$ is bounded by $d = \card{A}\cdot d_0 \leq p^{2k+1}$.

\medskip

Now assume that a $\cch{h}{p^k}$-circuit $\Gamma$ composes on the final level
$\cch{h-1}{p^k}$-circuits $\Gamma_1,\ldots,\Gamma_m$ by the gate $\cmod_{p^k}^{A}$.
To get the required polynomial $w_\Gamma(\o x)$ we simply plug into
$m$-ary $r_A(y_1,\ldots,y_m)$ the polynomials $w_{\Gamma_1},\dots,w_{\Gamma_m}$.
Obviously the degree of $w_\Gamma(\o x)$ is bounded by the maximal degree of the $w_{\Gamma_j}$'s
(i.e. by $d^{h-1}$) multiplied by the degree of $r_A(\o y)$ (i.e. by $d$)
which gives the required bound of $d^h$.
\hfill\myqed

\subsubsection*{Proof of Fact \ref{bar-poly}}

For an $n$-tuple of variables $\o x = (x_1,\ldots, x_n)$ we define
\[
v_j(\o x) = \sum_{1\leq i_1 < i_2 < \ldots <i_j \leq n} x_{i_1}\ldots x_{i_j}
\]
to be the sum of all $j$-linear monomials over the variables $\o x$.
In particular $v_0(\o x)=1$.
We will concentrate on their behaviour only for the boolean values $0,1$, so that
we put $v'_j : \set{0,1}^n \map \zp$ to be the appropriate restriction of $v_j$.

First observe that $v'_0,v'_1,\ldots,v'_n$
are linearly independent members of the vector space $\zp^{2^n}$.
Indeed, if $\sum_{j=0}^n \alpha_j v'_j = 0$
then evaluating at $\o x = (0,\ldots,0)$ we get $\alpha_0 =0$.
Moreover inducting on $j$ we evaluate on $\o x\in \set{0,1}^n$ with $1$ ocurring exactly $j$ times to get $\alpha_j=0$.

Now, fix $n \geq m = p^k$ and concentrate on the  $m$ dimensional subspace $V_m$ of the $2^n$ dimensional space $\zp^{2^n}$ spanned over $v'_0,\ldots,v'_{m-1}$.
One can easily see that each $v'_j$, and therefore each $v\in V_m$ is fully symmetric, i.e. $v(x_1,\ldots, x_n) = v(x_{\sigma 1},\ldots, x_{\sigma n})$ for all permutation $\sigma$.
This symmetry allows us to define $v[i]$ to be
$v(1,\ldots,1,0,\ldots,0)$ with exactly $i$ ones.

Slightly more effort is required to show that each $v'_j$ (and therefore each $v\in V_m$)
is $m$-periodic, i.e. $v[i+m]=v[i]$.
This reduces to show that $\binom{i+p^k}{j} =\binom{i}{j}$ modulo $p$,
or in other words,
that the prefixes of length $p^k$ of the $i$-th and $i+p^k$-th rows of Pascal triangle
coincide modulo $p$.
However due to the fact that an entry in the $j+1$-th row depends only on the two values in $j$-th  row, we are left with noticing that the mentioned coincidence holds for $i=0$,
or in other words that
\[
\binom{p^k}{j} \stackrel{p}{\equiv}
\left\{
\begin{array}{ll}
1, &\mbox{for $j=0$},\\
0, &\mbox{for $j=1,\ldots,p^k-1$}.
\end{array}
\right.
\]
To see that $V_m$ actually consists of all fully symmetric $m$-periodic functions
$\set{0,1}^n \map \zp$,
first note that each such a function can be obtained as a linear combination (over $\zp$)
of $w_0,w_1,\ldots,w_{m-1}$, where
\[
w_j[i] =
\left\{
\begin{array}{ll}
1, &\mbox{for $i \equiv j \mod p^k$},\\
0, &\mbox{else}.
\end{array}
\right.
\]
This shows that the vector space of all fully symmetric $m$-periodic functions
has dimension at most $m$ so that it has to coincide with $V_m$.

This observation allows us to represent the function $w' : \set{0,1}^n \map \zp$,
that behaves as $w$ in the statement of the Fact,
as a linear combination of the the $v'_j$'s.
This can be used to represent $w$ itself as the very same linear combination of the $v_j$'s
(with $j=0,1,\ldots,m-1$) showing that the degree of $w$ can be kept below $p^k$.
\hfill\myqed


\end{document}